\documentclass[11pt]{JHEP3}
\usepackage{amsmath,epsf,amssymb,latexsym,cite,graphics,setspace,bbm}
\usepackage[matrix,arrow,frame,import,curve,color]{xy}
\usepackage[active]{srcltx}

\DeclareFontFamily{U}{rsf}{}
\DeclareFontShape{U}{rsf}{m}{n}{
  <5> <6> rsfs5 <7> <8> <9> rsfs7 <10-> rsfs10}{}
\DeclareMathAlphabet\Scr{U}{rsf}{m}{n}

\def\C{{\mathbb C}}

\def\Aut{\operatorname{Aut}}

\def\im{\operatorname{im}}

\def\GL{\operatorname{GL}}

\def\GU{\operatorname{U{}}}

\def\p{\partial}

\def\la{\langle}
\def\ra{\rangle}

\def\cD{{\cal D}}
\def\cE{{\cal E}}

\def\cL{{\cal L}}

\def\cO{{\cal O}}

\def\cW{{\cal W}}

\newcommand\alphah{\widehat{\alpha}}

\newcommand\gammah{\widehat{\gamma}}
\newcommand\deltah{\widehat{\delta}}

\newcommand\kappah{\widehat{\kappa}}

\newcommand\sigmah{\widehat{\sigma}}

\newcommand\thetab{\overline{\theta}}

\newcommand\ah{\widehat{a}}

\newcommand\rh{\widehat{r}}

\newcommand\Qh{\widehat{Q}}

\def\cDb{\overline{\cD}}

\def\GUR{{\GU(1)_{\text{R}}}}
\def\GUL{{\GU(1)_{\text{L}}}}

\def\mon{{{\mathsf{M}}}}

\def\bdelta{{\boldsymbol{\delta}}}
\def\bdeltah{{\boldsymbol{\deltah}}}
\def\bgamma{{\boldsymbol{\gamma}}}
\def\bgammah{{\boldsymbol{\gammah}}}

\def\bsigma{{\boldsymbol{\sigma}}}
\def\bSigma{{\boldsymbol{\Sigma}}}
\def\bE{{\boldsymbol{E}}}
\def\be{{\boldsymbol{e}}}
\def\bH{{\boldsymbol{H}}}
\def\bsigmah{{\boldsymbol{\sigmah}}}

\def\lamr{{{\la m, \rho \ra}}}

\title{A (0,2) Mirror Map}
\author{Ilarion V.~Melnikov\\
\normalsize Max-Planck-Institut f\"ur Gravitationsphysik (Albert-Einstein-Institut),\\
\normalsize Am M\"uhlenberg 1, D-14476 Golm, Germany\\
Email:  \email{ilarion@aei.mpg.de}
}
\author{M.~Ronen Plesser\\
\normalsize Center for Geometry and Theoretical Physics, Box 90318\\
\normalsize Duke University, Durham, NC 27708-0318, USA\\
Email: \email{plesser@cgtp.duke.edu}
}

\abstract{ We study the linear sigma model subspace of the moduli space of (0,2) superconformal world-sheet theories obtained by deforming (2,2) theories based on Calabi-Yau hypersurfaces in reflexively plain toric varieties.  We describe a set of  algebraic coordinates on this subspace, formulate a (0,2) generalization of the monomial-divisor mirror map, and show that the map exchanges principal components of singular loci of the mirror half-twisted theories.  In non-reflexively plain examples the proposed map yields a mirror isomorphism between subfamilies of linear sigma models.  }

\preprint{AEI-2010-027}

\begin{document}

\section{Introduction}
The monomial-divisor mirror map (MDMM)~\cite{Aspinwall:1993rj} is an important tool in the study of (2,2) mirror symmetry.  In this work we will construct a generalization of this map that accounts for a class of (0,2) deformations of certain (2,2) theories.  This map should be a useful guide in explorations of the heterotic moduli space:  it may be used to efficiently determine singular loci of half-twisted theories, to compare topological heterotic rings on both sides of the mirror, and to have a hands-on algebraic description of the moduli space.  To describe our result, we will begin with a brief review of the MDMM in the (2,2) context.

 The geometric set-up is a Batyrev pair of mirror Calabi-Yau manifolds $(M,M^\circ)$, each constructed as a hypersurface in a $d$-dimensional Fano toric variety---$M\subset V$, $M^\circ \subset V^\circ$~\cite{Batyrev:1994hm}.  The crucial combinatorial ingredient in the construction is a $d$-dimensional reflexive lattice polytope $\Delta$.  This polytope plays a two-fold role in the mirror construction: on the one hand, it describes the Newton polytope for the hypersurface $P=0$ that defines $M \subset V$; on the other hand, the toric fan for $V^\circ$ is obtained by taking the cones over the faces of $\Delta$.  Thus, points in the faces of $\Delta$ have a dual interpretation:  they correspond to monomials in the polynomial defining $M\subset V$ and to divisors of $V^\circ$, which pull back to ``toric'' divisors on $M^\circ$.  This natural correspondence between the complex structure data for $M$ encoded in the coefficients of monomials and the (complexified) K\"ahler data encoded by the duals to the toric divisors can be used to construct an isomorphism between the ``polynomial'' and ``toric'' subspaces of deformations of $M, M^\circ$, respectively.  This is the monomial-divisor mirror map.

The MDMM has a physical realization in the context of gauged linear sigma models (GLSMs)---certain (2,2) supersymmetric abelian gauge theories that, for suitably tuned parameters, reduce at low energy to non-linear sigma models with target-spaces as above~\cite{Witten:1993yc}.  These theories are specified by the same combinatorial structure, and they depend on holomorphic parameters encoded in two types of superpotential terms.  The chiral superpotential for charged matter fields contains holomorphic couplings, while the twisted chiral superpotential contains the dependence on the Fayet-Iliopoulos and $\theta$-angle terms.  Not all of these parameters lead to deformations of the low energy SCFT, as some of them may be absorbed into irrelevant D-terms by appropriate field redefinitions; however, the combinatorial structure of the theory leads to a natural set of redefinition-invariant holomorphic parameters~\cite{Morrison:1994fr,Morrison:1995yh,Kreuzer:2010ph}.  Moreover, the coordinates so obtained, known as ``algebraic gauge'' coordinates, are exchanged by the MDMM and lead to an isomorphism between the chiral ring of the topological A-model and its B-model mirror~\cite{Morrison:1994fr,Morrison:1995yh,Batyrev:2002em,Szenes:2004mv,Borisov:2005hs,Karu:2005tr}.  These results suggest that the MDMM map between two families of linear sigma models leads to an isomorphism  of families of SCFTs in the low energy limit.

To describe a standard (0,2) non-linear sigma model (NLSM), in addition to choosing a target-space manifold $M$, we must also specify a holomorphic vector bundle $\cE \to M$.  The right-moving world-sheet fermions couple to the pullback of $T_M$, while the left-moving fermions couple to the pullback of $\cE$.  The theory has (2,2) SUSY provided that $\cE = T_M$.  As already noted in~\cite{Witten:1993yc}, the (2,2) GLSMs have a natural set of deformations that only preserve (0,2) SUSY; in a geometric phase, where the GLSM reduces to a NLSM at low energy, these correspond to certain unobstructed deformations of the tangent bundle.  We will refer to the space of models obtained via these deformations, together with already familiar K\"ahler and complex structure deformations, as the (0,2) GLSM moduli space.

Given a construction of mirror pairs at the (2,2) locus, it is natural to wonder whether the (0,2) GLSM moduli space for the pair $(M,\cE)$ is isomorphic to the GLSM moduli space for $(M^\circ,\cE^\circ)$.  In collaboration with M.~Kreuzer and J.~McOrist we investigated this question in~\cite{Kreuzer:2010ph}.  By considering the GLSM parameter space modulo (0,2) redefinitions, we described the (0,2) GLSM  moduli space, and  we found that for most mirror pairs the dimensions of the (0,2) GLSM moduli spaces did not agree.  However, we were able to identify a sufficient condition on the combinatorics of $\Delta$ and its polar dual $\Delta^\circ$ such that the dimensions matched:  no facet of $\Delta$ or $\Delta^\circ$ should contain a lattice point in its interior.  We called such geometries ``reflexively plain'' and found that there are roughly six million mirror families of three-folds with this property.

Of course simply matching the moduli space dimensions does not imply that there is any natural isomorphism between the theories.  In what follows, we will identify a natural isomorphism and show that it correctly maps an important physical property of the theory:  the isomorphism identifies the principal component of the singular locus (defined below) of an A/2-twisted GLSM with the principal component of the singular locus of its B/2-twisted mirror.  The half-twisted A/2 and B/2 theories, discussed in some detail in~\cite{McOrist:2008ji}, lead to computations of   $\boldsymbol{27}^3$ and $\boldsymbol{\overline{27}}^3$ unnormalized Yukawa couplings of charged matter fields in the $N=1$ space-time effective theory.  It is likely that the mirror map exchanges these Yukawa couplings as well, though we will not show it in this work.

In general the (2,2) GLSM moduli space is only a subspace of the full moduli space of complex structure and K\"ahler deformations:  certain variations of complex structure for $M \subset V$ cannot be represented by varying the coefficients of $P$, and some divisors on $V$ become reducible when restricted to $M$, thereby leading to K\"ahler deformations of $M$ not realized as hypersurfaces in $V$ .  Remarkably, the split in the moduli is preserved by mirror symmetry:  the space of toric K\"ahler deformations of $M$ is mirror to the space of polynomial deformations of $M^\circ$.  In the (0,2) context it is likely that bundle deformations also exhibit an analogue of  non-polynomial deformations---in the GLSM description these would be bundle deformations that cannot be realized by changing holomorphic couplings in the theory.  In models that are not reflexively plain such deformations may account for the discrepancy in the counting found in~\cite{Kreuzer:2010ph}.  In reflexively plain theories our results suggest that the GLSM moduli space forms a natural subspace in the full (0,2) moduli space, and this subspace is preserved by mirror symmetry.  

Although most powerful in the context of a reflexively plain mirror pair, the proposed mirror map also leads to mirror symmetric subfamilies in more general GLSMs.  Thus, while the full GLSM moduli space is not mirror symmetric, we can identify subspaces within the moduli space of $M$ and $M^\circ$ that are related by our mirror map.  The remaining GLSM parameters, which come from additional ``E-deformations,'' are more mysterious, at least from the point of view of the mirror map.  It should be borne in mind that they were typically not considered in work on stability of (0,2) GLSMs to instanton corrections~\cite{Silverstein:1995re,Basu:2003bq,Beasley:2003fx}.  In addition, the so-called ``non-linear'' E-deformations lead to significant technical problems in explicit computations in half-twisted theories~\cite{McOrist:2008ji}.  These issues certainly deserve further study, and we hope that our results on the mirror symmetric subfamilies will aid in such explorations.

The rest of the paper is organized as follows:  in section~\ref{s:setup} we will outline the structure of the (0,2) GLSMs under consideration; in section~\ref{s:coords} we will specialize to reflexively plain models, describe a natural set of coordinates on the GLSM moduli space, and present the conjecture for the (0,2) mirror map.  We will test the proposal by showing that it exchanges the singular loci of the mirror theories in~\ref{s:singular}.  Next, we extend the mirror map to subfamilies of more general GLSMs in~\ref{s:subfamilies}, and we end with a discussion of further directions and open questions.

\acknowledgments  It is a pleasure to thank P.~Aspinwall, J. McOrist, and E.~Miller for useful discussions.  The work of IVM is supported in part by the German-Israeli Project cooperation (DIP H.52) and the German-Israeli Fund (GIF).  MRP is supported in part by NSF grant DMS-0606578.

\section{The GLSM setup} \label{s:setup}
The (0,2) GLSM Lagrangian is most conveniently described in $(0,2)$ superspace with coordinates $x^-,x^+,\theta^+,\thetab^+$ and superspace covariant derivatives $\cD_+,\cDb_+$.  The field content is divided into chiral multiplets $Z_I, \Sigma_\alpha$; Fermi multiplets\footnote{These have left-moving fermions as lowest components.} $\Gamma^I$ satisfying
\begin{equation}
\cDb_+ \Gamma^I = E^I(Z,\Sigma);
\end{equation}
and vector fields $V_{-,a}, v_{+,a}$, $a=1,\ldots,r$.\footnote{We follow the notation of~\cite{Witten:1993yc}, and we stick to the set-up relevant to deformations of (2,2) theories for hypersurfaces in toric varieties.  Many generalizations are possible, e.g.~\cite{Distler:1995mi}.}
The field-strengths for the gauge fields transform  in gauge invariant chiral Fermi multiplets $\Upsilon_a$.  The ``matter'' fields $Z_I,\Gamma^I$ are charged under the gauge group with charges $Q^a_I$, while the $\Sigma_a$ multiplets are neutral.  A classical Lagrangian for the theory is given by
\begin{equation}
\cL = \cL_{\text{kin}} + \left\{ \int d\theta^+ \left[ \frac{\log(q_a)}{8\pi i} \Upsilon_a + \sum_I \Gamma^I \cW_I(Z) \right] + \text{h.c.}\right\},
\end{equation}
where $\{\cW_I(Z)\}$ is a set of holomorphic functions of the chiral multiplets with gauge charges $-Q^a_I$, while $q_a = e^{-2\pi r^a + i\theta^a}$ are holomorphic parameters combining the Fayet-Iliopoulos parameter and the theta angle for each gauge group.

To construct a GLSM for a Batyrev Calabi-Yau hypersurface, we pick a $d$-dimensional reflexive polytope $\Delta$ and call its dual polytope $\Delta^\circ$.  A reflexive polytope contains a unique interior point, and it is convenient to choose this to be the origin.  Since $\Delta,\Delta^\circ$ live in dual vector spaces, we can pair the lattice points in $\Delta$, labeled by $m$, with nonzero lattice points in $\Delta^\circ$, labeled by $\rho$.  In this way we define the rank $d$ pairing matrix $\la m,\rho \ra$.  
We describe the $d$-dimensional toric variety $V$ by using the homogeneous coordinate ring~\cite{Cox:1992bob, Cox:2000vi}.  For each $\rho$ we specify a complex coordinate $Z_{\rho}$ on $\C^{n}$ (i.e. $n$ is the number of non-zero lattice points in $\Delta^\circ$), and for a suitably chosen triangulation $\Sigma$ of $\Delta^\circ$, $V$ may be presented as a quotient
\begin{equation}
V = \frac{ \C^{n} - F_{\Sigma} }{G}.
\end{equation}
Here $G \simeq [\C^\ast]^{r}\times H$ for some finite abelian group $H$ , $r = n-d$, and $F_{\Sigma}$ is an exceptional set that depends on the chosen triangulation.  The action of $[\C^\ast]^r$ on the $Z_\rho$ is determined in terms of a basis for the kernel of $\lamr$.  We pick an integral basis $Q^a_{\rho}$, $a=1,\ldots,r$, and specify the action of $(t_1,\ldots,t_r)$ in $[\C^\ast]^r$ on $\C^n$ by
\begin{equation}
(t_1,\ldots,t_r) \cdot Z_\rho \mapsto \prod_{a} t_a^{Q^a_\rho} Z_\rho.
\end{equation}
This immediately leads to a homogeneous coordinate presentation of the defining polynomial $P$:
\begin{equation}
\label{eq:P}
P(Z) = \sum_{m \in \Delta} \alpha_m \mon_m, \quad \text{with}\quad \mon_m \equiv \prod_\rho Z_\rho^{\lamr + 1} ,
\end{equation}
for some choice of parameters $\alpha_m$.\footnote{The sum over $m$ includes $m=0$ here and below unless otherwise indicated.}
$P(Z)$ transforms homogeneously under the action of $G$, with charges $-Q_0^a \equiv \sum_\rho Q^a_\rho.$  

With these combinatorial ingredients in hand, we can describe the (2,2) GLSM for $M \subset V$ in terms of (0,2) fields.
We take the matter field index $I$ to run over the lattice points in $\Delta^\circ$, and we set the gauge charges of the fields to be $Q^a_\rho$.  Finally, we take $r$ gauge-neutral matter fields $\Sigma_\alpha$ and consider the action specified by
\begin{align}
E^{\rho} = \sum_{\alpha=1}^r \Sigma_\alpha Q^\alpha_\rho Z_\rho, \quad \Gamma^I\cW_I = \Gamma^0 P(Z) + Z_0 \sum_\rho \Gamma^\rho \frac{\p P}{\p Z_{\rho}}.
\end{align}
This theory has (2,2) supersymmetry, as well as unbroken (and non-anomalous) $\GUL\times\GUR$ R-symmetries.  The model depends on the parameters $q_a$, $\alpha_m$, which in a geometric phase give rise to, respectively, complexified K\"ahler and complex structure deformations. 

  To obtain (0,2) deformations, we simply generalize the superpotential couplings while preserving gauge invariance and $\GUL\times\GUR$ symmetries.\footnote{In the low energy limit $\GUR$ should correspond  to the $R$-symmetry of the (0,2) theory, while $\GUL$ should become an important left-moving symmetry~\cite{Distler:1995mi}.}  At this point it is useful to combine the $r$ $\Sigma_\alpha$ into a vector $\bSigma$, which allows us to express the allowed couplings as
\begin{align}
\label{eq:Edefsgen}
E^0 = Z_0 \bSigma\cdot \bdelta, \quad E^{\rho} =  \bSigma\cdot \bE^{\rho}(Z),
\end{align}
where $\bdelta$ is a vector of parameters and $\bE^\rho(Z)$ is a vector of polynomials with same gauge charges as $Z_\rho$.
The $\Gamma^I \cW_I$ terms take the form
\begin{align}
\sum_I \Gamma^I \cW_I = \Gamma^0 P(Z) + Z_0 \sum_\rho \Gamma^\rho J_\rho(Z),
\end{align}
where each $J_\rho$ has the same set of monomials as $P_{,\rho}$,  but with coefficients unrelated to the $\alpha_m$ in $P$.
Since the Fermi multiplets are not chiral, supersymmetry of the holomorphic superpotential is not automatic.  Demanding (0,2) SUSY leads to
constraints on the parameters:
\begin{equation}
\label{eq:susyconstgen}
 P(Z) \bdelta + \sum_\rho  J_\rho(Z) \bE^{\rho}(Z)= 0 \quad \text{for all Z}.
\end{equation}

A desire for brevity and a weariness of repetition have led to much interesting toric geometry and physics being left out from the preceding discussion.  Before we move on to study reflexively plain models, we will comment on a few of these points.  First, the finite abelian group $H$ must be included as a discrete gauge group of the GLSM, leading to an orbifold of the theory.  Second, we have not commented on the exceptional set $F_{\Sigma}$, partly because there are many possible exceptional sets, each corresponding to a ``phase'' of the GLSM---with different phases corresponding to different cones in the space of Fayet-Iliopoulos parameters. Existence of certain triangulations known as maximal projective subdivisions show that there exist $F_\Sigma$---equivalently a phase of the GLSM---such that $M \subset V$ has suitably mild singularities.  Finally, the toric description of the automorphism group of the toric variety plays a crucial role in describing the general form of $\bE^{\rho}$ and the various field redefinitions that may be used to eliminate some of the holomorphic parameters.  For models with (2,2) supersymmetry the discussion of these redefinitions follows the original construction of the MDMM, while in the (0,2) models these details are discussed at length in~\cite{Kreuzer:2010ph}. 

\section{GLSM moduli space of reflexively plain models}\label{s:coords}
We now restrict the general set-up to the reflexively plain models.  A reflexively plain pair of polytopes leads to toric varieties $V,V^\circ$ with smallest possible automorphism groups---namely those where the continuous automorphisms are inherited from $[\C^\ast]^d$ reparametrizations of the algebraic torus contained in $V (V^\circ)$.  The form of $P(Z)$ and $J_\rho(Z)$ remains the same as in general GLSMs, with $P(Z)$ given in eqn.~(\ref{eq:P}), and $J_\rho$ determined by
\begin{align}
\label{eq:J}
Z_\rho J_\rho &= \sum_{m\in\Delta} j_{m\rho}  \mon_m,
\end{align}
where the parameters $j_{m\rho}$ obey 
\begin{equation}
j_{m\rho} = 0\quad\text{whenever}\quad \lamr = -1.
\end{equation}
On the (2,2) locus these are given by $j_{m\rho} = \alpha_m (\lamr+1)$.

The $E$ couplings simplify for a reflexively plain model:
\begin{equation}
\label{eq:Edefsplain}
\bE^{\rho}(Z) = \be^{\rho} Z_\rho,
\end{equation}
with $\be^{\rho}$ a vector of field-independent parameters.   The holomorphic field re-definitions are also considerably simpler to describe:
\begin{equation}
Z_I \mapsto u_I Z_I, \quad \Gamma^I \mapsto v_I \Gamma^I,
\end{equation}
where $u_I,v_I \in \C^\ast$.  In addition, there are $\GL(r,\C)$ rotations of the $\Sigma_\alpha$ multiplets, which we will choose to write as
\begin{equation}
\bSigma \mapsto u_0^{-1} v_0   \bSigma\cdot\bH
\end{equation}
for $\bH\in\GL(r,\C)$.
The SUSY constraint requires
\begin{equation}
\alpha_m\bdelta + \sum_\rho  j_{m\rho} \be^\rho= 0 \quad\text{for all}~m.
\end{equation}

The redefinitions induce an action on the parameters, and two sets of parameters related by such a rescaling are expected to lead to identical low energy physics.  The action on the $\bdelta,\be^\rho$, $\alpha_m$, and $j_{m\rho}$ is easy to determine:
\begin{align}
\be^\rho  &\mapsto \frac{u_\rho v_\rho^{-1}}{u_0 v_0^{-1}} \bH \be^\rho ,\qquad
\bdelta \mapsto  \bH\bdelta , \nonumber\\
\alpha_m &\mapsto \alpha_m\times v_0 \prod_\rho u_\rho^{\la m,\rho\ra +1} , \nonumber\\
j_{ m\rho} & \mapsto j_{m\rho} \times u_0 v_\rho u_\rho^{\la m,\rho \ra} \prod_{\rho'\neq\rho} u_\rho^{\la m,\rho'\ra+1}.
\end{align}
The redefinitions also act on the complexified K\"ahler parameters $q_a$.  The origin of this action is that the fermion measure transforms anomalously under transformations with $u_\rho \neq v_\rho$.  Holomorphy allows us to determine this shift by considering the $\GU(1)$ anomalies for transformations with $|u_\rho| = |v_\rho| = 1$. It is not hard to see that under such a transformation the action shifts by 
\begin{equation}
\Delta \cL = \sum_a \int d\theta^+ \frac{1}{8\pi i}  \log\left[\prod_\rho \left(\frac{u_\rho v_\rho^{-1}}{u_0 v_0^{-1}}\right)^{Q^a_\rho}\right] \Upsilon_a,
\end{equation}
whence we conclude that the redefinition leads to
\begin{equation}
q_a \mapsto q_a \times \prod_{\rho} \left[ \frac{u_\rho v_\rho^{-1}}{u_0 v_0^{-1}}\right]^{Q^a_\rho}.
\end{equation}

The parametrization of the GLSM just described is analogous to homogeneous coordinates for a projective space, and it has proven useful in arguments for stability of (0,2) deformations~\cite{Silverstein:1995re}.  For our purposes, it will be more convenient to have an affine set of coordinates independent of the redefinitions.  To construct these, we first note that the action on the  $\alpha_m$ is the same as in the (2,2) case, and there is a well-known way to pick invariant coordinates~\cite{Aspinwall:1993rj,Morrison:1994fr}.  We introduce the integral rank $d$ pairing matrix $\pi_{m\rho} = \lamr$ for $m\neq 0$ and choose an integral basis $\Qh^{\ah}_m$ for its cokernel.\footnote{The $\Qh^{\ah}_m$, together with $\Qh^{\ah}_0 = -\sum_{m\neq 0} \Qh^{\ah}_m$, will be the gauge charges for the mirror GLSM.  Of course $Q^a_\rho$ are an integral basis for the kernel of $\pi_{m\rho}.$}  It is then easy to see that 
\begin{equation}
\kappah_{\ah} \equiv \prod_{m\neq 0} \left( \frac{\alpha_m}{\alpha_0} \right)^{\Qh^{\ah}_m}
\end{equation}
are invariant under the redefinitions.  Next, we introduce
\begin{equation}
\kappa_a  \equiv q_a \prod_\rho \left(\frac{j_{0\rho}}{\alpha_0}\right)^{Q^a_\rho}
\end{equation}
as invariant ``K\"ahler'' coordinates.  Note that on the (2,2) locus the $\kappa_a$ reduce to the usual $q_a$.  We also define 
\begin{equation}
\bgamma^\rho \equiv \frac{j_{0\rho}}{\alpha_0} \be^\rho, \quad\text{and}
\quad b_{m\rho} \equiv \frac{\alpha_0 j_{m\rho}}{\alpha_m j_{0\rho}}-1\quad\text{for}\quad m\neq 0.
\end{equation}
The motivation for the latter choice is that on the (2,2) locus $b_{m\rho} = \pi_{m\rho}$.  Note that we have assumed  $\alpha_m\neq 0$.  This does not constitute much of a loss of generality and will simplify a number of arguments.

We now have a set of invariant parameters $b_{m\rho}, \kappah_{\ah}, \kappa_a$, as well the $\bdelta,\bgamma$ that transform to $\bH \bdelta,\bH\bgamma$.  These are not quite independent because of the SUSY constraint, which takes the form
\begin{align}
\alpha_0 \bdelta + \sum_\rho \alpha_0 \bgamma^\rho & = 0, \nonumber\\
\alpha_m \bdelta + \sum_\rho \alpha_m (b_{m\rho}+1)\bgamma^\rho &= 0, \quad\text{for}\quad m\neq 0.
\end{align}
With our assumption $\alpha_m\neq 0$, we find that these take an elegant form:
\begin{equation}
\bdelta = - \sum_\rho \bgamma^\rho, \quad  \sum_\rho b_{m\rho}\bgamma^\rho = 0 \quad \text{for} \quad m \neq 0.
\end{equation}
At the (2,2) locus the matrix $b_{m\rho}$ has rank $d$, which determines the $\bgamma^\rho$ to be in the $r$-dimensional kernel of $b$ spanned by the $Q^a_\rho$.  Thus, the $\bgamma^\rho$ are determined up to the $\GL(r,\C)$---the remaining field redefinitions.  In fact, in order to obtain a non-singular theory the rank of $b_{m\rho}$ can be at most $d$.  Otherwise, at least one of the components of $\bgamma^{\rho}$ must necessarily vanish, leading to an unconstrained $\Sigma$ multiplet.  In fact, as we will argue in the next section, the theory will also be singular if the rank of $b_{m\rho}$ drops below $d$.  Thus, we will restrict attention to $b_{m\rho}$ of rank $d$, which implies that $\bgamma^\rho$ and $\bdelta$ are determined precisely up to the $\GL(r,\C)$ redundancy in a choice of basis for the kernel of $b_{m\rho}$.

We are now left with ``affine'' coordinates $\kappa_a$, $\kappah_{\ah}$, as well as a rank $d$ matrix $b_{m\rho}$ that satisfies
\begin{equation}
b_{m\rho} = -1  \quad \text{whenever} \quad \pi_{m\rho} = -1.
\end{equation}
Do these give the expected dimension of the GLSM moduli space?  Our deformation parameters are easily counted in terms of $\ell(\Delta), \ell(\Delta^\circ)$---the numbers of lattice points in $\Delta$ and $\Delta^\circ$, as well as $K$---the number of pairs $m,\rho$ with $\pi_{m\rho} = -1$:
\begin{align}
\# (\kappa )  & = \ell(\Delta^\circ) -1 -d, \nonumber\\
\# (\kappah) & = \ell(\Delta) -1 -d , \nonumber\\
\# (b )           & = d\left[ \ell(\Delta) +\ell(\Delta^\circ) - 2 -d \right] -K,
\end{align}
where in the last line we used the fact that a $p\times q$ matrix of rank at most $d$ is an irreducible subvariety of codimension $(p-d)(q-d)$ in the space of $p\times q$ matrices~\cite{Harris:book}.  Adding up these parameters, we find the dimension of the GLSM moduli space for a reflexively redundant theory~\cite{Kreuzer:2010ph}.  We note that the (0,2) deformations encoded by $b_{m\rho}$ have an elegant algebraic description as an intersection of   $K$ linear equations with the moduli space of matrices of rank at most $d$.

The parametrization just constructed suggests a simple conjecture for the mirror theory: \emph{the mirror GLSM is obtained by exchanging $\Delta, \Delta^\circ$, transposing the matrix $b$, and exchanging the roles of $\kappa_a$ and $\kappah_{\ah}$.}

\section{The singular loci} \label{s:singular}
The reader may agree that mirror map conjectured above is elegant enough to be easily believable; however, it would be nice to have a more convincing test.  Perhaps the simplest convincing test is to check that the map identifies the singular loci of the A/2 and B/2 models with their mirrors.  In this section we will see that this is indeed so.

\subsection{The A/2 singular locus}
The singular locus of the A/2 theory is easily determined by using the Coulomb branch techniques introduced in~\cite{Witten:1993yc,Morrison:1994fr} and extended to (0,2) theories in~\cite{McOrist:2007kp,McOrist:2008ji}.  More precisely, this identifies a number of components of the singular locus associated to divergences due to non-compact $\Sigma_\alpha$ directions.  The so-called ``principal component'' of the singular locus---a complex co-dimension subvariety in the moduli space has a particularly simple form as the locus where the $r$ equations
\begin{equation}
\prod_\rho \left[ \frac{\bsigma\cdot \be^\rho}{\bsigma\cdot\bdelta}\right]^{Q^a_\rho} = q_a, 
\end{equation}
have a solution for some $\bsigma \neq 0$.  This is nicely rewritten in terms of our invariant coordinates as
\begin{equation}
\label{eq:A2locus}
\prod_\rho \left[ \frac{\bsigma\cdot \bgamma^\rho}{\bsigma\cdot\bdelta}\right]^{Q^a_\rho} = \kappa_a.
\end{equation}

On the $(2,2)$ locus this reduces to the familiar condition~\cite{Morrison:1994fr}
\begin{equation}
\label{eq:Alocus}
\prod_I \left(\sum_{a=1}^r \sigma_a Q_I^a\right)^{Q^a_I} = q_a.
\end{equation}
Another interesting limit is obtained by working deep in a geometric phase, where quantum effects may be neglected, and the vacuum expectation value of $Z_0$ is zero.    In this case, a solution to eqn.~(\ref{eq:A2locus}) implies that the classical theory develops a flat $\bsigma$ direction.  It is not hard to see from the classical Lagrangian that this implies there exists $\bsigma \neq 0$ such that
\begin{equation}
\label{eq:Aclassloc}
\bsigma\cdot \bgamma^\rho Z^\ast_\rho = 0 \quad\text{for some}\quad Z^\ast \in V .
\end{equation}
 The geometric import of this singularity is easy to understand.  The deformed bundle is defined by the complex\footnote{Here the $D_\rho$ are the toric divisors, and $\cO(D_\rho)|_M$ are the corresponding line bundles restricted to $M$.}
\begin{equation}
\label{eq:bundledef}
\xymatrix{ 0\ar[r] &\cO^{r}|_M \ar[r]^-{e^{\alpha\rho} Z_\rho} &\oplus_\rho\cO(D_\rho)|_{M}  \ar[r]^-{J_\rho} & \cO(\sum_\rho D_\rho)|_{M}\ar[r] & 0 },
\end{equation}
with space of sections of $\cE$ described by $\ker J / \im E$.  
Given a solution to eqn.~(\ref{eq:Aclassloc}),  $\dim \im E(Z^\ast) < r$, and the supersymmetry constraint implies $Z^\ast \in M$. The rank of $\cE$ increases at $Z^\ast$, and, consequently, the sheaf $\cE$ is no longer a bundle.

Thus, the (0,2) singular locus in eqn.~(\ref{eq:A2locus}) interpolates between the (2,2) singular locus of eqn.~(\ref{eq:Alocus}), and the classical singular locus of eqn.~(\ref{eq:Aclassloc}).  It is amusing to note~\cite{McOrist:2008ji} that a classically singular bundle can lead to a perfectly sensible theory away from the large radius limit.

\subsection{The B/2 singular locus}
The principal component of the B/2 singular locus requires a little bit more work.  Experience from the (2,2) locus and the results of~\cite{McOrist:2008ji} suggest that these singularities are due to a non-compact direction for the $Z_0$ multiplet. Before we embark on our computation, we should note that there are a number of remaining issues in the B/2 model:  for example, there is no complete proof that it is independent of the $q_a$.  It is easy to see that ``most'' instantons do not contribute to B/2 correlators, but a full non-renormalization argument has yet to be given. We will assume this independence does indeed hold, which allows us to work in a geometric phase. 

The $Z_0$ multiplet develops a non-compact direction if and only if there exists a point $p \in M$ such that $J_\rho(p) = 0$ for all $\rho$.  In fact, without loss of generality we may work with $p \in V$, since the SUSY constraint ensures that $P(p) = 0$ will follow from $J_\rho(p) = 0$.  We will concentrate on the situation when $p \in [\C^\ast]^d\subset V$, which in the case of (2,2) theories yields the principal component of the singular locus.\footnote{The point $p$ may also belong to some $[\C^\ast]^k$, $k < d$ in the compactification of $[\C^\ast]^d$ to $V$; the resulting singular locus may either be a limit of the principal component, or give rise to a new component.  In the latter case, in (2,2) models there is a well-defined combinatorial description of these additional components on both sides of the mirror.}    We will continue to use the same terminology in the (0,2) theories.

We seek the conditions on the $b_{m\rho}$ and $\alpha_m$ such that $Z_\rho J_\rho (p) = 0$ for all $\rho$ for some $p\in [\C^\ast]^d$.  In the case of (2,2) models, where $b_{m\rho} = \pi_{m\rho}$, a characterization of the principal component of the singular locus was obtained by Kapranov~\cite{Kapranov:1991}.  What follows is a simple generalization of his result.  We start by recasting $Z_\rho J_\rho$ in terms of invariant parameters:
\begin{equation}
Z_\rho J_\rho = \frac{ j_{0\rho} }{\alpha_0} \left[ \alpha_0\mon_0+ \sum_{m\neq 0} \alpha_m \mon_m (b_{m\rho}+1) \right].
\end{equation}
Thus, to find a singularity we must solve 
\begin{equation}
\alpha_0\mon_0+ \sum_{m\neq 0} \alpha_m \mon_m (b_{m\rho}+1) = 0 \quad\text{for all} \quad \rho.
\end{equation}
The problem is very easy to solve if we consider the $\mon_m$ as independent variables.  If we choose a basis $\{\gammah^{\alphah m}\}$ for the cokernel of $b_{m\rho}$, the general solution depends on a set of parameters $\sigmah_{\alphah}$  and takes the form
\begin{equation}
\alpha_m \mon_m = \bsigmah \cdot \bgammah^m, \quad\text{for}~m\neq 0\quad\text{and}\quad \alpha_0 \mon_0 = \bsigmah\cdot \bdeltah,
\end{equation}
where $\bdeltah = -\sum_{m\neq 0} \bgammah^m.$  Of course the $\ell(\Delta)$ terms $\alpha_m \mon_m$ are not independent but instead for any $Z$ satisfy $\rh = \ell(\Delta)-d-1$ constraints
\begin{equation}
\prod_{m\neq 0} \left[ \frac{\alpha_m \mon_m}{\alpha_0 \mon_0} \right]^{\Qh^{\ah}_m} = \prod_{m\neq 0}\left[ \frac{\alpha_m}{\alpha_0} \right]^{\Qh^{\ah}_m} = \kappah_{\ah},
\end{equation}
where as before the $\Qh^{\ah}_m$ constitute a basis for the cokernel of $\pi_{m\rho}$.  Plugging in the result of the linear problem, we see that a singular point exists only if
\begin{equation}
\label{eq:B2locus}
\prod_{m\neq 0} \left[ \frac{\bsigmah\cdot \bgammah^m}{\bsigmah\cdot\bdeltah}\right]^{\Qh^{\ah}_m} = \kappah_{\ah}, 
\end{equation}
for some non-zero $\bsigmah$.  In fact, it is not hard to see that the converse is also true:  a solution to eqn.~(\ref{eq:B2locus}) guarantees that we can find a singular point.  This is most easily seen by taking a logarithm of $\alpha_m \mon_m/\alpha_0 \mon_0$, which leads to simpler equations for $Z_\rho$:
\begin{equation}
\sum_\rho \pi_{m\rho} \log Z_\rho = \log \frac{\alpha_0 \bsigmah\cdot \bgammah_m}{\alpha_m \bsigmah\cdot \bdeltah}.
\end{equation}
By eqn.~\ref{eq:B2locus} the right-hand side is in the image of $\pi_{m\rho}$, and we will be able to find a requisite $Z_\rho$.  Thus, the B/2 singular locus  is described as the locus of solutions to eqn.~\ref{eq:B2locus} for some non-zero $\bsigmah$.  This yields $\rh$ equations for the $\sigmah_{\alphah}$; however, invariance of the equations under rescaling $\bsigmah$ implies that a solution will be found on a complex codimension one subvariety in the moduli space.   

The B/2 singular locus, like its A/2 cousin, interpolates between two familiar notions: singularities of $M$ and singularities of $\cE$. On the (2,2) locus, where $J_\rho = P_{,\rho}$, a solution to eqn.~(\ref{eq:B2locus}) guarantees a singularity in the complex structure of $M$; even when $M$ is a non-singular hypersurface, a look at the complex in eqn.(\ref{eq:bundledef}) shows that a simultaneous solution to $J_\rho = 0$ for some $Z^\ast \in M$ implies that the rank of $\cE$ increases as $Z^\ast$, leading to a singular bundle.  Of course there is an important difference between the two singular loci:  while the B/2 locus is entirely determined by classical computations, the A/2 locus encodes quantum singularities, which in a geometric phase correspond to diverging instanton sums.

Inspection of the principal components of the A/2 locus in eqn.~(\ref{eq:A2locus}) and the B/2 locus in eqn.~(\ref{eq:B2locus}) makes it clear that the proposed mirror map will map these to, respectively, the B/2 and A/2 singular loci of the mirror.  This provides a non-trivial check of the conjecture.

It should be easy to extend this correspondence to other components of the singular loci.  On the A/2 side these are associated to mixed Higgs/Coulomb singularities~\cite{Morrison:1994fr}, while on the B/2 side they are due to singularities that occur at points where some of the $Z_\rho$ vanish.

The discussion of the B/2 loci assumed that $b_{m\rho}$ has rank $d$; when the rank drops below $d$, the theory must inevitably become singular, since the existence of extra $\sigmah_{\alphah}$ in eqn.~(\ref{eq:B2locus}) will mean solutions for any values of the parameters.  When interpreted from the mirror A/2 side, the appearance of an extra $\sigmah_{\alphah}$ suggests that this singularity may be due to a new branch in the moduli space, which is perhaps more easily interpreted in a different GLSM.

\section{Mirror Subfamilies}\label{s:subfamilies}
In the previous sections we constructed a set of coordinates for the GLSM moduli space of reflexively plain geometries, suggested a generalization of the monomial-divisor mirror map, and showed that the generalization is consistent with the form of the singular loci of the theory.  While this is a nice result, it is so far restricted to a relatively small class of Calabi-Yau hypersurfaces.  However, as we will now explain, the idea can be fruitfully applied to generic GLSMs for Calabi-Yau hypersurfaces, where it yields mirror symmetric subfamilies in the full GLSM moduli space.

As found in~\cite{Kreuzer:2010ph}, the difficulty of matching the GLSM parameters with those of its mirror lies in the more complicated form of the E-deformations and field redefinitions.  The combinatoric source of the trouble is the possible existence of lattice points contained in the interiors of facets of $\Delta$ and $\Delta^\circ$. We will let $W(W^\circ)$ be the number of such points in $\Delta(\Delta^\circ)$.  On the (2,2) locus the lattice interior facet points in $\Delta$ give rise to $W$ redundant complex structure parameters, while the $W^\circ$ points in $\Delta^\circ$ give rise to redundant K\"ahler parameters.  The former arise from coefficients in $P$ that may be eliminated by elements of $\Aut V$ that do not belong to $[\C^\ast]^d$; the latter correspond to divisors of $V$ that do not intersect the hypersurface $M\subset V$.  Although these are indeed redundant, the MDMM naturally extends to a map on these parameters as well~\cite{Morrison:1994fr}, and in some sense the map is simpler to state if we allow for the redundancy.  The point is that as long as the redundancy is understood on both sides of the mirror, it does not really lead to difficulties.

To extend this idea to the (0,2) case, we require an additional step:  we must restrict the set of E-deformations of eqn.~(\ref{eq:Edefsgen}) to just the ``diagonal'' form of eqn.~(\ref{eq:Edefsplain}).  Having done this, we can construct the parameters invariant under the ``diagonal'' redefinitions exactly as above, both for the model and its mirror.  The two sets of parameters will be exchanged by the mirror map exactly as in the reflexively plain examples, yielding a mirror pair of subfamilies, each of dimension
\begin{equation}
N_{\text{diag}} = (d+1) (\ell(\Delta)+\ell(\Delta^\circ) -2 -d) -d -K,
\end{equation}
where $K$ is the number of pairs $m,\rho$ with $\pi_{m\rho} = -1$.  As on the $(2,2)$ locus, we expect $W+W^\circ$ of these parameters to be redundant; happily, this redundancy is itself mirror symmetric.  Moreover, since we have an explicit mapping of the coordinates on our subspace, it is easy (as with the MDMM) to specialize the construction to subfamilies.  

\section{Discussion}\label{s:discussion}
A conjecture for the (0,2) mirror map allows us to pursue a number of new lines of inquiry.  Perhaps the most obvious direction would be to prove that the map yields an isomorphism of topological heterotic rings~\cite{Adams:2003zy, Adams:2005tc} by developing a generalization of the toric residue methods that were used in~\cite{Morrison:1994fr,Batyrev:2002em,Szenes:2004mv,Borisov:2005hs,Karu:2005tr}.  There is one immediate issue that one must confront, since, unlike the topological field theories associated to (2,2) theories, the half-twisted models do not have a clear relationship between local observables and deformations of the action.  Experience with (2,2) models and the form of the (0,2) mirror map do suggest a guess for how to map the correlators.  The natural observables of the B/2-twisted theory for $M\subset V$ consist of monomials $\cO_m =  \alpha_m Z_0\mon_m$, while the mirror A/2-twisted observables are given by the $\cO^\circ_m = \bgammah^m \cdot \bsigmah$.  It is tempting to suggest that the mirror map should exchange these via
\begin{equation}
\la \cO_{m_1} \cO_{m_2} \cO_{m_3} \ra_{\text{B/2,}M} = \la \cO^\circ_{m_1} \cO^\circ_{m_2} \cO^\circ_{m_3} \ra_{\text{A/2},M^\circ}.
\end{equation}
Is this true without a parameter-dependent change of basis?  If not, can a basis change lead to an equivalence?

The ``non-diagonal'' E-deformations continue to pose a challenge.  We have found a way to avoid them, either by working with models without such complications, or working on certain subspaces of the moduli space.  However, this is not entirely satisfactory.  Are these deformations perhaps lifted by world-sheet instantons?  If not, how do we describe their mirrors? Having answered these questions one would have a reasonably complete picture of what the GLSM may teach us about (0,2) theories with a (2,2) locus.

While developing this complete picture is surely important and likely to lead to interesting mathematical structures and neat physical effects, a larger world awaits!  Many rich (0,2) theories without a (2,2) locus can be studied by GLSM techniques~\cite{Witten:1993yc,Distler:1993mk}; while comparatively little is known about them away from special points in the moduli space, there are intriguing hints of mirror pairs and a rich duality structure~\cite{Distler:1995bc,Blumenhagen:1996vu}.  It would be very useful to have an analogue of a mirror map for at least some theories in this large class.  There will be new difficulties, but we believe at least some of the interplay between combinatorics and physics should work in a familiar fashion.   Our work offers a small but sturdy stepping stone into the ``real'' (0,2) world.

\bibliographystyle{utphys}
\providecommand{\href}[2]{#2}\begingroup\raggedright\endgroup

\end{document}